\documentclass[12pt]{article}
\usepackage{setspace}
\usepackage{epsfig}
\usepackage{amssymb}
\def\be{\begin{equation}}
\def\ee{\end{equation}}
\def\ba{\begin{array}}
\def\ea{\end{array}}
\def\beqn{\begin{eqnarray}}
\def\eeqn{\end{eqnarray}}

\def\bt{\begin{tabular}}
\def\et{\end{tabular}}
\def\bc{\begin{center}}
\def\ec{\end{center}}

\begin{document}
\title{Revisiting texture 5 zero quark mass matrices}
\author{Priyanka Fakay \\ \textit{Department of Physics, D.A.V. College, Sector-10,
 Chandigarh-160011} \\ \textit{Department of Physics, Panjab University, Sector-14,
 Chandigarh-160014}\\
 \textit{priyanka.fakay@gmail.com}}

\maketitle \abstract{ The question of viability of texture 5 zero
Fritzsch-like quark mass matrices are examined in the context of
the latest quark mixing data} \vskip 1.0cm

           It has been shown that all possible combinations of texture 6
     zero quark mass matrices, Fritzsch-like as well as non Fritzsch-like are already ruled out \cite{singreview},\cite{neelu}.
     Similarly, all possible combinations of hermitian texture 5 zero quark mass matrices are also
      ruled out except for the case when $M_U$
     is texture 3 zero type and $M_D$ is texture 2 zero type \cite{neelu}.
     The purpose of
     the present work is to examine the
     implications of recently refined data on texture specific quark mass
      matrices.  The plan of paper is as follows. In Section I we
     discuss the essentials of formalism of texture 5 zero quark
     mass matrices and in Section II we discuss the inputs used
     for analyses.In Section III we present  discussion of analyses
     and in Section IV we summarize our conclusions.

\section{Texture specific mass matrix and CKM matrix \label{3qfor}}
For the sake of readability, we underline some of the essentials
of the methodology leading to the construction of the $V_{CKM}$.
To define the various texture specific cases considered here, we
present the typical Fritzsch like texture specific hermitian quark
mass matrices, for example,
 \be
 M_{U}=\left( \ba{ccc}
0 & A _{U} & 0      \\ A_{U}^{*} & D_{U} &  B_{U}     \\
 0 &     B_{U}^{*}  &  C_{U} \ea \right), \qquad
M_{D}=\left( \ba{ccc} 0 & A _{D} & 0      \\ A_{D}^{*} & D_{D} &
B_{D}     \\
 0 &     B_{D}^{*}  &  C_{D} \ea \right),
\label{nf2zero}\ee
 where $M_{U}$ and $M_{D}$ correspond to up and
down mass matrices respectively, $A$ and $B$ are complex elements
and $C$ is a real element of the mass matrix. It may be noted that
each of the above matrix is texture 2 zero type with $A_{k}
=|A_k|e^{i\alpha_k}$
 and $B_{k} = |B_k|e^{i\beta_k}$, where $k= U,D$.
The texture specific Fritzsch-like mass matrices
 discussed above, can be exactly diagonalized and the
corresponding $V_{ckm}$ can be constructed from these.
 To illustrate the procedure as well as to facilitate
discussion, we detail  the construction of $V_{ckm}$. Further, we
define $\phi_1 =  \alpha_U- \alpha_D$, $\phi_2= \beta_U-
 \beta_D$. The complex matrix $M_i$ ($i=U,D$) can be expressed as $ M_i=
P_i^{\dagger} M_i^r P_i \,,  $ where $ P_i= {\rm diag}
(e^{-i\alpha_i},\,1,\,e^{i\beta_i})\ $ and the real matrix $M_i^r$
is \be M_i^r = \left( \ba  {ccc} 0 & |A_i| & 0 \\ |A_i| & D_i &
|B_i|
\\
            0 & |B_i| & C_i \ea \right) \,. \ee
The $M_i^r$ can be diagonalized by the orthogonal
  transformation, for example,
\be M_i^{\rm diag} = O_i^T M_i^{r} O_i \equiv
  O_i^T P_i M_i P_i^{\dagger} O_i
 \,,   \label{o1}\ee
where \be M_i^{\rm diag} = {\rm diag}(m_1,\,-m_2,\,m_3)\,, \ee
 the subscripts 1, 2 and 3 refer respectively to $u,\, c$ and $t$
  for the $U$ sector and $d,\,s$ and $b$ for the $D$ sector. It
  may be noted that the second mass eigen value is chosen with a
  negative sign to facilitate the construction of the
  diagonalizing transformation $O_i$, for the details in this
  regard we refer the reader to \cite{singreview}.

Using the invariants, tr$M_i^r$,  tr${M_i^r}^2$ and det$M_i^r$,
the values of the matrix elements $|A_i|$, $|B_i|$ and $C_i$ can
be expressed in terms of quark masses as \beqn
  C_i& = &(m_1-m_2+m_3-D_i)\,, \\
   |A_i| &=&(m_1 m_2 m_3/C_i)^{1/2}\,, \\
 |B_i| &= &
 [(m_3-m_2-D_i)(m_3+m_1-D_i)(m_2-m_1+D_i)/C_i]^{1/2}\,.
 \eeqn

 The details of the relationship between the mass matrices and the
  mixing matrix are given \cite{singreview}, for example,
\be V_{\rm CKM} = O_{U}^{T} P_{U} P_{D}^{\dagger} O_{D} =
V_{U}^{\dagger} V_{D}\,,  \label{v} \ee where the unitary matrices
$V_{U}( = P^\dagger_U \, O_U$) and $V_{D}( = P^\dagger_D \, O_D$)
are the diagonalizing  matrices for the hermitian matrices $M_U$
and $M_D$ respectively.

\section{Inputs used for analysis \label{inputs5}}
 For ready
reference as well as to facilitate the discussion, the input quark
masses
  at the $m_Z$  scale \cite{xing2012} used in the analysis are given as
 \beqn m_u = 1.27^{+0.50}_{-0.42}\, {\rm MeV},~~~~~m_d =
2.90^{+1.24}_{-1.19}\, {\rm MeV},~~~~ m_s=55^{+16}_{-15}\, {\rm
MeV},~~~~~~~~\nonumber\\ ~~~~~m_c=0.619 ^{+0.084}_{-0.084}\, {\rm
GeV},~~ m_b=2.89 ^{+0.09}_{-0.09}\, {\rm GeV},~~ m_t=171.1
^{+3.0}_{-3.0} \, {\rm GeV}. ~~~\label{qmasses} \eeqn
 The light quark masses $m_u$, $m_d$ and $m_s$ have been
further constrained by using the  mass ratios \cite{leut}.
 For the purpose of our calculations, we are giving full variation to
phases $\phi_{1}$ and $\phi_{2}$ from 0 to $2\pi$. We have assumed
the condition of  naturalness of mass matrices implying that $D_U$
and $D_D$   be restricted in the ranges (0.0464-0.4870) $m_t$ and
(0.0276-0.4448) $m_b$ respectively. While carrying out the analyses,
we  first attempt to reproduce $V_{us}$, $V_{ub}$ and $V_{cb}$
corresponding respectively to $s_{12}$, $s_{23}$ and $s_{13}$ as
well as CP asymmetry parameter Sin2$\beta$. It may be mentioned that
for our analysis, we have considered a wide range of $V_{ub} =
(4.15\pm0.49)\times 10^{-3}$, which includes both its exclusive and
inclusive values.
\section{Results and discussion \label{rd}}

To begin with, we first discuss the two cases of texture 5 zero
Fritzsch-like mass matrices which can be obtained from equation
(\ref{nf2zero}) by taking either $D_{U}$ = 0 and $D_{D} \neq 0$ or
$D_{U} \neq 0$ and $D_{D}$ = 0. To check the viability of these 2
cases of texture 5 zero quark mass matrices, we attempt to reproduce
the CKM matrix. However, we find out that out of the two cases, only
one possibility $D_U$=0 case has limited viability, while the other
case  seems to be ruled out. For this purpose, by giving full
variation to phases $\phi_1$ and $\phi_2$ as well as to
$D_U$ or $D_{D}$ and by including the constraints of only $|V_{us}|$
and $|V_{ub}|$, we have made an attempt to calculate the CKM matrix
element $V_{td}$.
 \begin{figure}[hbt]
\begin{minipage}{0.45\linewidth}   \centering
\includegraphics[width=2.in,angle=-90]{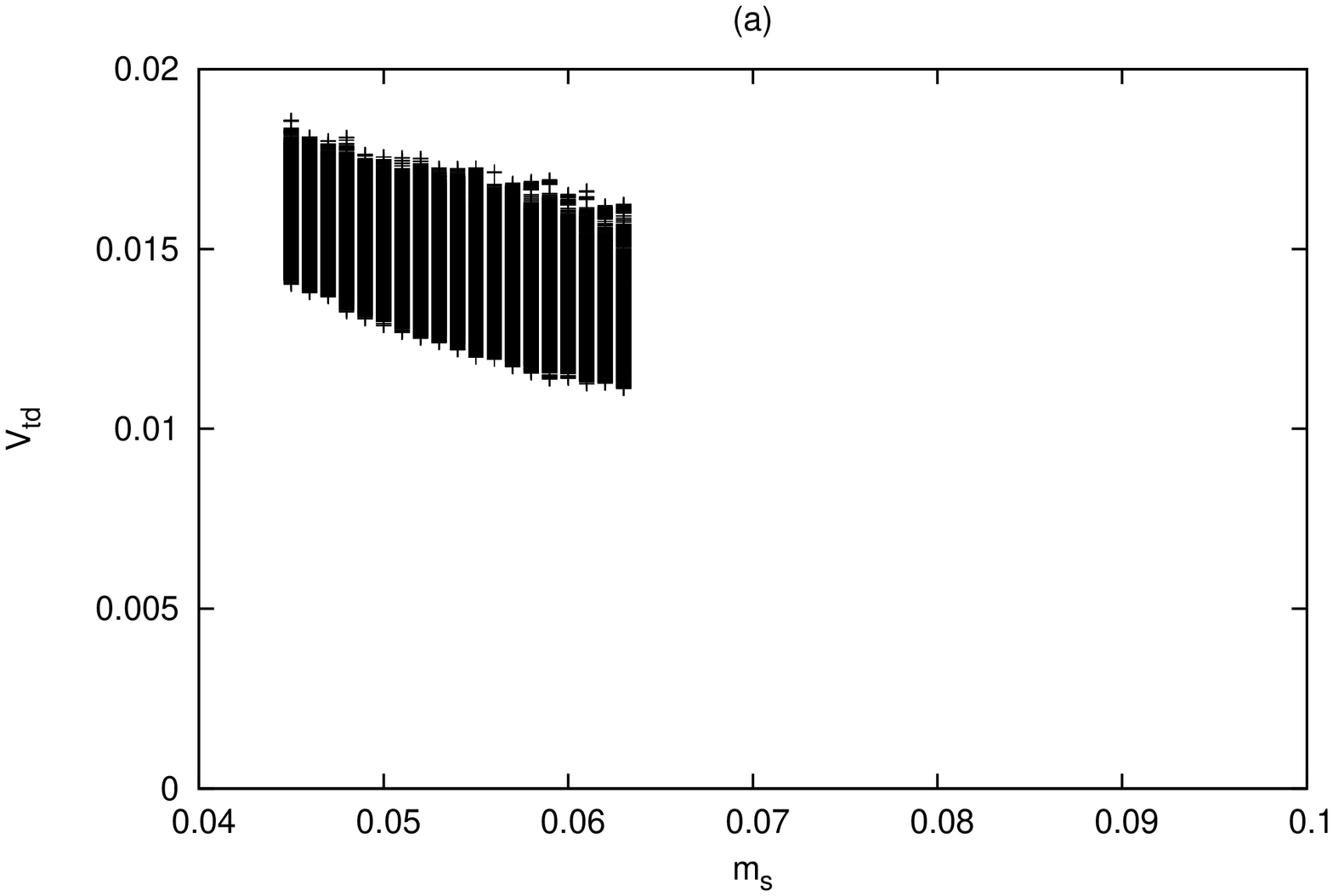}
    \end{minipage} \hspace{0.5cm}
\begin{minipage} {0.45\linewidth} \centering
\includegraphics[width=2.in,angle=-90]{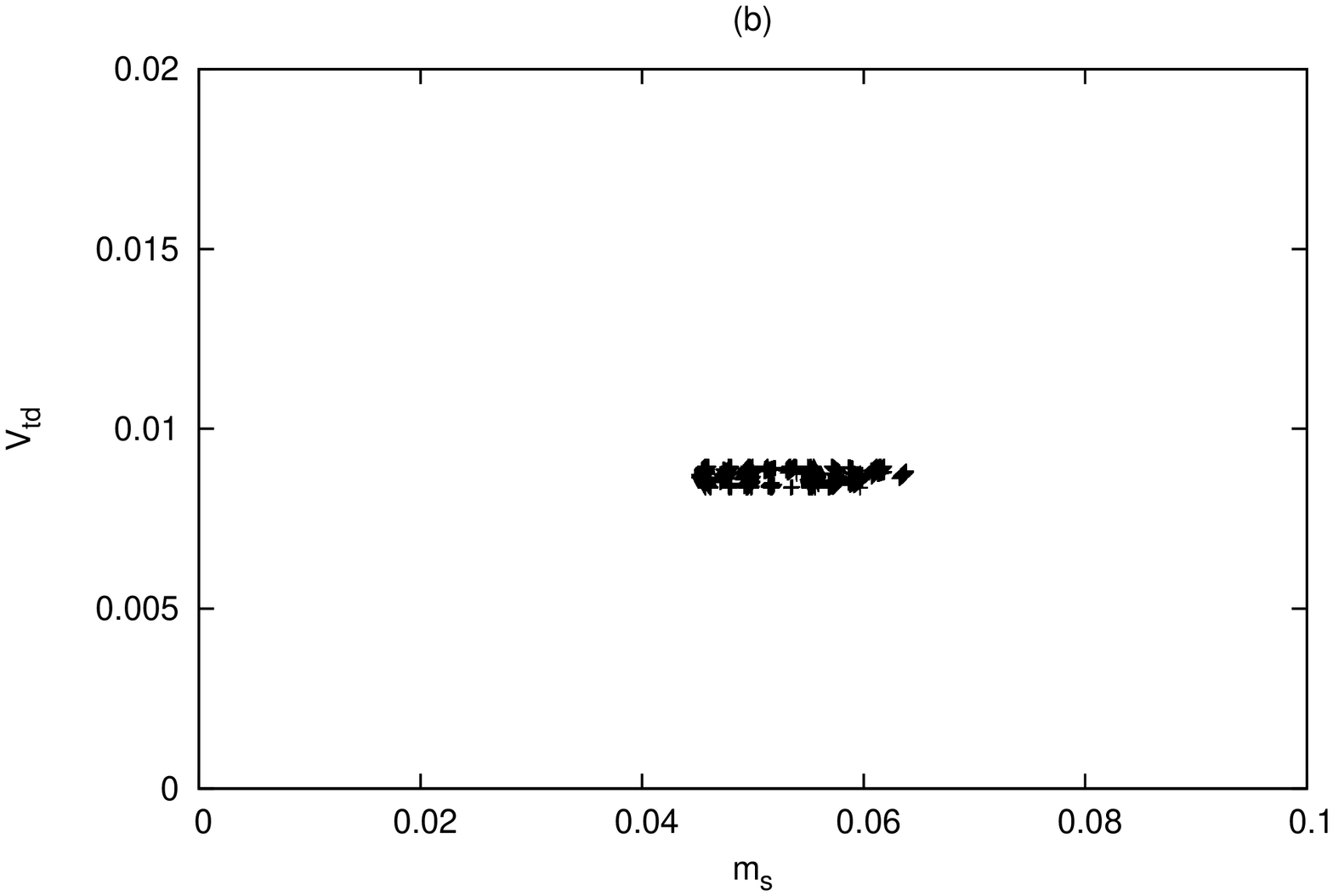}
  \end{minipage}\hspace{0.5cm}
   \caption{Plot
showing variation of $V_{td}$ with the strange quark mass $m_s$
 for the two cases of texture 5 quark zero mass matrices (a)$D_D$=0 case (b) $D_U$=0 case.}
  \label{fig15}
  \end{figure} To this end, in Figures (\ref{fig15}a) and (\ref{fig15}b) we have plotted
the CKM matrix element $V_{td}$ against the strange quark mass for $D_D$=0 and $D_U$=0 case of
 texture 5 zero mass matrices respectively. From Figure
(\ref{fig15}a) it is immediately clear that for all values of $m_s$
considered here, the range of $V_{td}$ is much higher than the
values quoted by PDG \cite{pdg}, whereas, from (\ref{fig15}b) one
finds that $V_{td}$ can be obtained within the allowed range.
However, in case the higher values of $m_s$ than the one considered
here are used then this case is also ruled out.

 \section{Summary and conclusions}
To summarize, we have analyzed texture 5  zero
      quark mass matrices keeping in mind the recently refined
      data. Using the latest inputs regarding masses and mixing
            parameters, we find that the texture 5 zero $D_D$=0
             and $D_U\neq 0$ case is completely ruled out whereas
             the other texture 5 zero $D_U$=0
             and $D_D\neq 0$ case has limited viability depending
             upon the mass of strange quark $m_s$ used.

             \textbf{Acknowledgements} \vskip0.05 cm   P.F. would like to
acknowledge Principal D.A.V. College and  the Chairperson,
Department of Physics, P.U., for providing facilities to work.
P.F.  would also like to acknowledge  M. Gupta for useful
discussions.

 \end{document}